\documentclass[runningheads]{llncs}
\usepackage{multirow}
\usepackage{soul}

\usepackage[colorlinks=true,linkcolor=blue]{hyperref}
\usepackage{graphicx}

\usepackage{epsfig}
\usepackage{amsmath}
\usepackage{amssymb}
\usepackage[capitalize]{cleveref}
\crefname{section}{Sec.}{Secs.}
\Crefname{section}{Section}{Sections}
\Crefname{table}{Table}{Tables}
\crefname{table}{Tab.}{Tabs.}

\usepackage{xcolor}         
\usepackage{pifont}

\definecolor{myred}{RGB}{220,50,47} 
\definecolor{mygreen}{RGB}{133,153,0}
\definecolor{commentcolor}{RGB}{133,153,0}
\definecolor{urlcolor}{rgb}{0.93,0.01,0.55}

\newcommand{\diffrecon}{DMCVR}

\newcommand{\etal}{\textit{et al}. }

\newcommand{\method}[1]{#1}
\newcommand{\important}[1]{#1}

\newcommand{\lastday}[1]{#1}

\begin{document}
%
\title{DMCVR: Morphology-Guided Diffusion Model for 3D Cardiac Volume Reconstruction}
%
\titlerunning{DMCVR: Diffusion Model for 3D Cardiac Volume Reconstruction}
%


\author{Xiaoxiao He\inst{1}\orcidID{0000-0003-4581-0712} \and Chaowei Tan\inst{2} \and Ligong Han\inst{1} \and Bo Liu\inst{3} \and Leon Axel\inst{4} \and Kang Li\inst{5} \and Dimitris N. Metaxas\inst{1}\thanks{Corresponding Author}}
%
\authorrunning{X. He et al.}
\institute{Department of Computer Science, Rutgers University \and FocusAI Inc. \and
Walmart Global Tech\and School of Medicine, New York University\and 
West China Biomedical Big Data Center, Sichuan University West China Hospital}
\maketitle              
\begin{abstract}
Accurate 3D cardiac reconstruction from cine magnetic resonance imaging (cMRI) is crucial for improved cardiovascular disease diagnosis and understanding of the heart's motion. However, current cardiac MRI-based reconstruction technology used in clinical settings is 2D with limited through-plane resolution, resulting in low-quality reconstructed cardiac volumes. 
\method{To better reconstruct 3D cardiac volumes from sparse 2D image stacks, we propose \important{a morphology-guided diffusion model for 3D cardiac volume reconstruction, \diffrecon{}, that synthesizes high-resolution 2D images and corresponding 3D reconstructed volumes. Our method outperforms previous approaches by conditioning the cardiac morphology on the generative model, eliminating the time-consuming iterative optimization process of the latent code, and improving generation quality.}} The learned latent spaces provide global semantics, local cardiac morphology and details of each 2D cMRI slice with highly interpretable value to reconstruct 3D cardiac shape. 
Our experiments show that \diffrecon{} is highly effective in several aspects, such as 2D generation and 3D reconstruction performance. With \diffrecon{}, we can produce high-resolution 3D cardiac MRI reconstructions, surpassing current techniques. Our proposed framework has great potential for improving the accuracy of cardiac disease diagnosis and treatment planning. Code can be accessed at \url{https://github.com/hexiaoxiao-cs/DMCVR}.

\end{abstract}
%
%
%

\section{Introduction}
\begin{figure}[t]
    \centering    \includegraphics[width=0.95\linewidth]{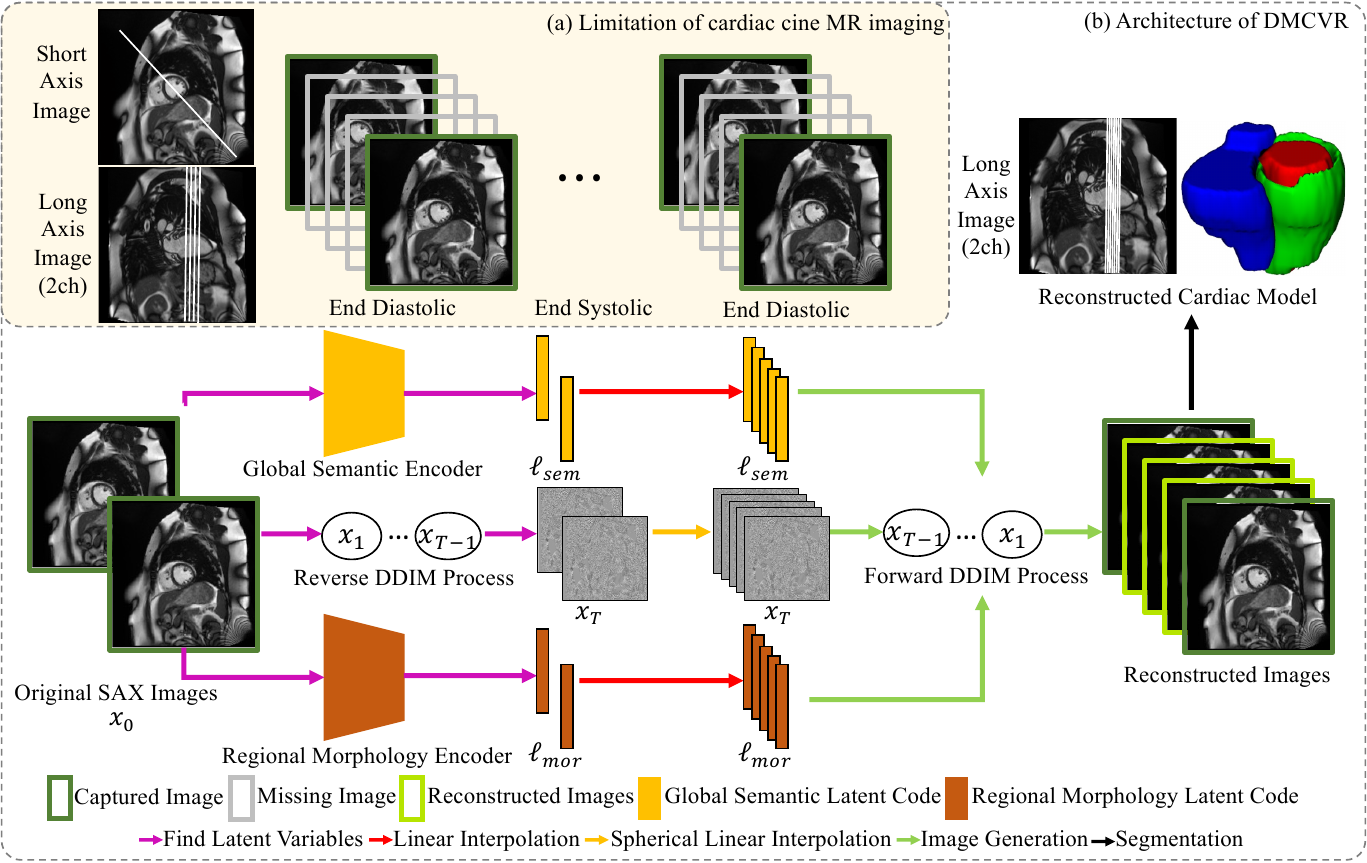}
    \caption{(a) demonstrates the limitations of cardiac cMRI. The white line in the short axis (SAX) image is the location of 2 chamber (2ch) long axis (LAX) image slice and vice versa. The grey images indicate the missing slices which are not captured during the MRI scan. (b) is an overview of our \diffrecon{} architecture. 
    The SAX images $x_0$ are first encoded to global semantic $\ell_{sem}$, regional morphology $\ell_{mor}$ and stochastic latent codes $x_T$, followed by interpolation in their respective latent space. The reconstructed images are sampled from a forward denoising diffusion implicit model (DDIM) process conditioned on the three latent codes. Finally, the 3D cardiac model is reconstructed via stacking the labels. The red, green, and blue regions represent the left ventricle cavity (LVC), left ventricle myocardium (LVM), and right ventricle cavity (RVC), respectively.}
    \label{fig:diffrecon}
\end{figure}

Medical imaging technology has revolutionized the field of cardiac disease diagnosis, enabling the assessment of both cardiac anatomical structures and motion, including the creation of 3D models of the heart \cite{van1999quantification}. 
Cardiac cine magnetic resonance imaging (cMRI) \cite{pelc1991phase,sechtem1987quantification} is widely used in clinical diagnosis \cite{patel2016diagnostic}, allowing for non-invasive visualization of the heart in motion with detailed information on cardiac function and anatomy \cite{peng2016review}. While cMRI has great potential in helping doctors understand and analyze cardiac function \cite{isensee2018automatic,pattynama1994evaluation}, the imaging technique has certain drawbacks including low through-plane resolution to accommodate for the limited scanning time, as visualized in \cref{fig:diffrecon}. 
\lastday{Recently, researchers have approached the problem of cardiac volume reconstruction with learning-based generative models \cite{chang2022deeprecon}. However, most of the methods suffer from low generation quality, missing key cardiac structures and long generation times. This paper focuses on improving the cardiac model generation quality, while reducing the generation time, aiming to better reconstruct the missing structure of the cardiac model from low through-plane resolution cMRI.}

Conventional 3D cardiac modeling \cite{lopez2015three} consists of 2D cardiac image segmentation followed by 3D cardiac volume reconstruction. 
Recent advances in deep learning methods have shown great success in medical image segmentation \cite{gao2022data,he2019effective,liu2022transfusion,zhangli2022region}. 
\method{After obtaining 2D labels, the neighboring labels are stacked to reconstruct the 3D model. Nevertheless, due to the low inter-slice spatial cMRI resolution, a significant amount of structural information is lost in the resulting 3D volume. Thus, the interpolation between cMRI slices is necessary. }
\lastday{Traditional intensity-based interpolation methods 
often yield blurring effects and unrealistic results. 
Conventional deformable model-based method \cite{myronenko2010point} does not need consistency across images of the corresponding cardiac structures, but requires image-based structure segmentation which is nontrivial and hinders their ability to generalize.} 
To overcome these limitations, an end-to-end pipeline based on generative adversarial networks (GANs), DeepRecon, was recently proposed in \cite{chang2022deeprecon} that utilizes the latent space to interpolate the missing information between adjacent 2D slices. \method{The generative network is first trained and a semantic image embedding in the $\mathcal{W}^+$ space \cite{abdal2019image2stylegan} is computed. Evidently, the acquired semantic latent code is not optimal and needs iterative optimization with segmentation information for improving image qualities.} 
\method{However, even with the optimization step, the generated images still miss details in the cardiac region, which indicates the $\mathcal{W}^+$ space DeepRecon found does not represent the heart accurately.}

In order to eliminate the step for optimizing the latent code and improve the image generation quality, we propose a morphology-guided diffusion-based 3D cardiac volume reconstruction method that improves the axial resolution of 2D cMRIs through global semantic and regional morphology latent code interpolation as indicated in \cref{fig:diffrecon}. \method{Inspired by \cite{preechakul2022diffusion}, we utilize the global semantic latent code to encode the image into a high-level meaningful representation of the image. \lastday{To improve the cardiac volume reconstruction, our approach needs to focus on the cardiac region. Therefore, we introduce the regional morphology latent code which represents the shapes and locations of LVC, LVM and RVC, which will help generating the cardiac region.}} The method consists of three parts: an implicit diffusion model, a global semantic encoder and a segmentation network that encodes an image to regional morphology embeddings. The proposed method does not require iteratively fine-tuning the latent codes. Our contributions are: 1) the first diffusion-based method for 3D cardiac volume reconstruction, 2) introducing the local morphology-based latent code for improved conditioning on the image generation process, 3) $8\%$ improvement of left ventricle myocardium (LVM) segmentation accuracy and $35\%$ improvement of structural similarity index compared to previous methods, and  4) improved efficiency by eliminating the iterative step for optimizing the latent code.
%
\section{Methods}
\begin{figure}[t]
    \centering
    \includegraphics[width=0.8\linewidth]{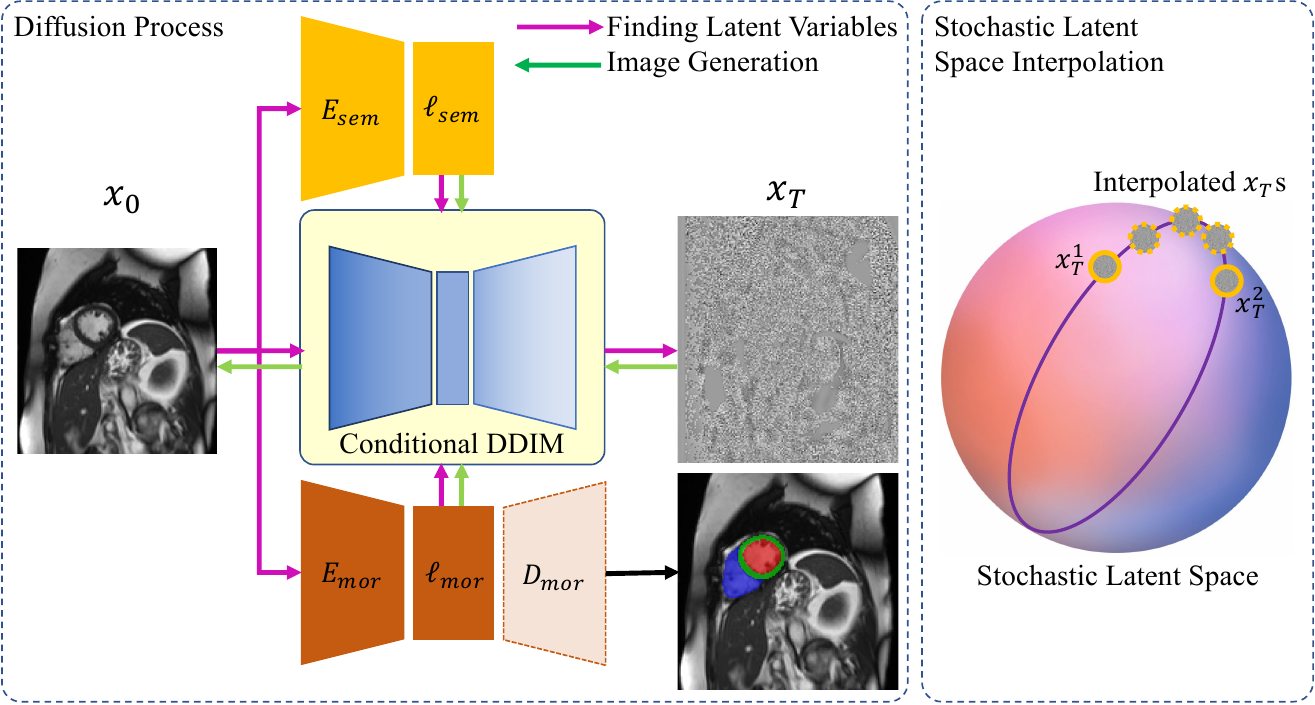}
    \caption{On the left side, we demonstrates the network structure of the \diffrecon{}, which consists of an global semantic encoder, a regional morphology encoder/decoder and a conditional DDIM. The right side shows the visualization of the stochastic latent space sampled from a high-dimensional Gaussian distribution $\mathcal{N}(0,I)$.}
    \label{fig:network}
\end{figure}
\cref{fig:network} demonstrates the structure of our \diffrecon{} approach that learns the global semantic, regional morphology, and stochastic latent spaces from MR images to yield a broad range of outcomes, including generation of high-quality 2D image and high-resolution 3D reconstructed volume. In this section, we will first describe the architecture of our \diffrecon{} method and then elaborate on the latent space-based 3D volume generation which enables 3D volume reconstruction.

\subsection{\diffrecon{} Architecture}
Our \diffrecon{} is composed of a global semantic encoder $E_{sem}$, a regional moprphology network $(E_{mor},D_{mor})$ and a diffusion-based generator $G$. The generating process $G$ is defined as follows: given input $x_T,\ell_{sem},\ell_{mor}$, which are the stochastic, global semantic and regional morphology latent codes, we want to reconstruct the image $x_0$ recursively as follows:
\begin{equation}
    x_{t-1}=\sqrt{\alpha_{t-1}}f_\theta(x_t,t,\ell_{sem},\ell_{mor})+\sqrt{1-\alpha_{t-1}}\epsilon_\theta(x_t,t,\ell_{sem},\ell_{mor}),
\end{equation}
where $\epsilon_\theta(x_t,t,\ell_{sem},\ell_{mor})$ is the noise prediction network and $f_\theta$ is defined as removing the noise from $x_t$ or Tweedie’s formula \cite{efron2011tweedie}:
\begin{equation}
    f_\theta(x_t,t,\ell_{sem},\ell_{mor})=\frac{1}{\sqrt{\alpha_t}}(x_t-\sqrt{1-\alpha_t}\epsilon_\theta(x_t,t,\ell_{sem},\ell_{mor}))
\end{equation}
Here, the term $\alpha_t$ is a function of $t$ affecting the sampling quality.

The forward diffusion process takes the noise $x_T$ as input and produces $x_0$ the target image. Since the change in $x_T$ will affect the details of the output images, we can treat $x_T$ as the stochastic latent code. Therefore, finding the correct stochastic latent code is crucial for generating image details. Thanks to DDIM proposed by Song \etal \cite{song2021denoising}, it is possible to get $x_T$ in a deterministic fashion by running the generative process backwards to obtain the stochastic latent code $x_T$ for a given image $x_0$. This process is viewed as a stochastic encoder $x_T=E_{sto}(x_0,\ell_{sem},\ell_{mor})$, which is conditioned on $\ell_{sem}$ and $\ell_{mor}$. \lastday{This conditioning helps us to remove the iterative optimization step used by previous method. We formulate the inversion process from $x_0$ to  $x_T$ as follows:}
\begin{equation}
  x_{t+1}=\sqrt{\alpha_{t+1}}f_\theta(x_t,t,\ell_{sem},\ell_{mor})+\sqrt{1-\alpha_{t+1}}\epsilon_\theta(x_t,t,\ell_{sem},\ell_{mor})
\end{equation}

Although using the stochastic latent variables we are able to reconstruct the image accurately, the stochastic latent space does not contain interpolatable high-level semantics. Here we utilize a semantic encoder proposed by Preechakul \etal \cite{preechakul2022diffusion} to encode the global high-level semantics into a descriptive vector for conditioning the diffusion process, similar to the style vector in StyleGAN \cite{karras2019style}. The global semantic encoder utilizes the first half of the UNet, and is trained end-to-end with the conditional diffusion model. 

\method{One drawback of the global semantic encoder is that it encodes the general high-level features, but tends to pay little attention to the cardiac region. This is due to the relatively small area of LVC, LVM and RVC in the cMRI slice. 
However, the generation accuracy of the cardiac region is crucial for the cardiac reconstruction task. For this reason, we introduce the regional morphology encoder $E_{mor}$ that embeds the image into the latent space containing necessary information to produce the segmentation map of the target cardiac tissues. With this extra morphology information, we are able to guide the generative model to focus on the boundary of the ventricular cavity and myocardium region, which will  produce increased image accuracy in the cardiac region and the downstream segmentation task. Here, we do not assume any particular architecture for the segmentation network. However, in our experiments, we utilize the segmentation network MedFormer proposed by Gao \etal \cite{gao2022data} for its excellent performance.}

The training of \diffrecon{} contains the training of the segmentation network and the training of the generative model. We first train the segmentation model with summation of focal loss and dice loss \cite{gao2022data}.
We utilize the simple loss introduced in \cite{ho2020denoising} for training the conditional diffusion implicit model, where
\begin{equation}
    \label{eqn:norm}
    L_{gen}(x)=\mathbb{E}_{t\sim \text{Unif}(1,T),\epsilon\sim \mathcal{N}(0,I)}||\epsilon_\theta(x_t,t,E_{sem}(x_0),E_{mor}(x_0))-\epsilon||^2_2.
\end{equation}

\subsection{3D volume reconstruction and latent-space-based interpolation}
Due to various limitations, the gap between consecutive cardiac slices in cMRI is large, which results in an under-sampled 3D model. In order to output a smooth super-resolution cine image volume, we generate the missing slices by using the interpolated global semantic, regional morphology and stochastic latent codes. For global semantic and regional morphology latent code $\ell$, since it is similar to the idea of latent code in StyleGAN, we utilize the same interpolation strategies as in the original paper between adjacent slices. Assume that $k<j-i, i<j$,
\begin{equation}
    \ell^{i+k}=(1-\frac{k}{j-i})\ell^i+\frac{k}{j-i}\ell^j.
\end{equation}

For interpolating the stochastic latent variable, it is important to consider that the distribution of stochastic noise is high-dimensional Gaussian, as shown in \cref{eqn:norm}. Thus, our stochastic embedding is positioned on a sphere shown in \cref{fig:network}. Using linear interpolation on the stochastic noise deviates from the underlying distribution assumption and causes the diffusion model to generate unrealistic images. Hence, to preserve the Gaussian property of the stochastic latent space, we interpolate the stochastic latent codes over a unit sphere, which can be written as follows: Let $k<j-i, i<j$ and $x_T^i\cdot x_T^j = \cos{\theta}$,

\begin{equation}
    x_T^{i+k}=\frac{\sin((1-\frac{k}{j-i})\theta)}{\sin(\theta)}x_T^i+\frac{\sin(\frac{k}{j-i}\theta)}{\sin(\theta)}x_T^j.
\end{equation}
\section{Experiments}
\subsection{Experimental Settings}
In this study we use data from the publicly available UK Biobank cardiac MRI data \cite{petersen2015uk}, which contains SAX and LAX cine CMR images of normal subjects. LVC, LVM and RVC 
are manually annotated on SAX images at the end-diastolic (ED) and end-systolic (ES) cardiac phases. We use 808 cases containing 484,800 2D SAX MR slices for training and 200 cases containing 120,000 2D images for testing. To evaluate the 3D volume reconstruction performance, we randomly choose 50 testing 2D LAX cases to evaluate the 3D volume reconstruction task. All models are implemented on PyTorch 1.13 and trained with 4$\times$RTX8000. 

\subsection{Evaluation of the 2D slice generation quality}
\begin{table}[t]
    \centering
    \caption{Quantitative comparison among the segmentation results of the original image (Original), DeepRecon with 1k optimization steps (DeepRecon$_{1k}$), Diffusion AutoEncoder~\cite{preechakul2022diffusion} (DiffAE) and our \diffrecon{}. We use a pretrained segmentation model on images generated by different methods. All metrics are evaluated against the ground truth based on 3D SAX images.}
    \begin{tabular}{c|c|c|c|c|c|c}
    \hline
Cardiac Region & Method & DICE $\uparrow$ & VOE$\downarrow$ & ASD$\downarrow$ & HD$\downarrow$ & ASSD$\downarrow$ \\
 \hline
\multirow{4}{*}{All labels} & Original & \textit{0.943} & \textit{10.730} & \textit{0.229} & \textit{4.056} & \textit{0.229}\\
\cline{2-7}
& DeepRecon$_{1k}$ & 0.914 & 15.179 &  0.367 & 5.879 & 0.397 \\ 
\cline{2-7}
 & DiffAE & 0.919 & 14.913  & 0.322 & 4.654 & 0.326 \\
\cline{2-7}

 & \diffrecon{} & \textbf{0.935} & \textbf{12.153}  & \textbf{0.261} & \textbf{4.093} &\textbf{0.266} \\
 \hline

\multirow{4}{*}{LVC} & Original &\textit{0.937} & \textit{11.579}  & \textit{0.221} & \textit{3.156} & \textit{0.224}\\ \cline{2-7}
& DeepRecon$_{1k}$ & 0.928 & 12.955  & 0.336 & 4.299 & 0.328 \\
\cline{2-7}
 & DiffAE & 0.910 & 16.049 & 0.330 & 3.710 & 0.320 \\
 \cline{2-7}
 & \diffrecon{} & \textbf{0.929} & \textbf{12.940}  & \textbf{0.250} & \textbf{3.236} & \textbf{0.254} \\
 \hline
\multirow{4}{*}{LVM} & Original & \textit{0.875} & \textit{22.082}  & \textit{0.226} & \textit{3.140} & \textit{0.237}\\
\cline{2-7}
& DeepRecon$_{1k}$ & 0.796 & 33.382  & 0.390 & 5.730 & 0.389 \\
\cline{2-7}
 & DiffAE & 0.825 & 29.333 & 0.351 & 4.032 & 0.338 \\
 \cline{2-7}
 & \diffrecon{} & \textbf{0.865} & \textbf{23.636} & \textbf{0.282} & \textbf{3.519}& \textbf{0.267} \\
 \hline
\multirow{4}{*}{RVC} & Original & \textit{0.898} & \textit{18.187} & \textit{0.273} & \textit{4.458} & \textit{0.267}\\
\cline{2-7}
& DeepRecon$_{1k}$ & 0.858 & 23.662  & 0.381 & 6.304 & 0.473 \\
\cline{2-7}
 & DiffAE & 0.857 & 24.518 & 0.346 & 5.217 & 0.382 \\
 \cline{2-7}
 & \diffrecon{} & \textbf{0.884} & \textbf{20.467}  & \textbf{0.273} & \textbf{4.460} & \textbf{0.308} \\
 \hline
    \end{tabular}
    \label{tab:seg-comparison}
\end{table}
\begin{figure}[t]
    \centering
    \includegraphics[width=0.8\linewidth]{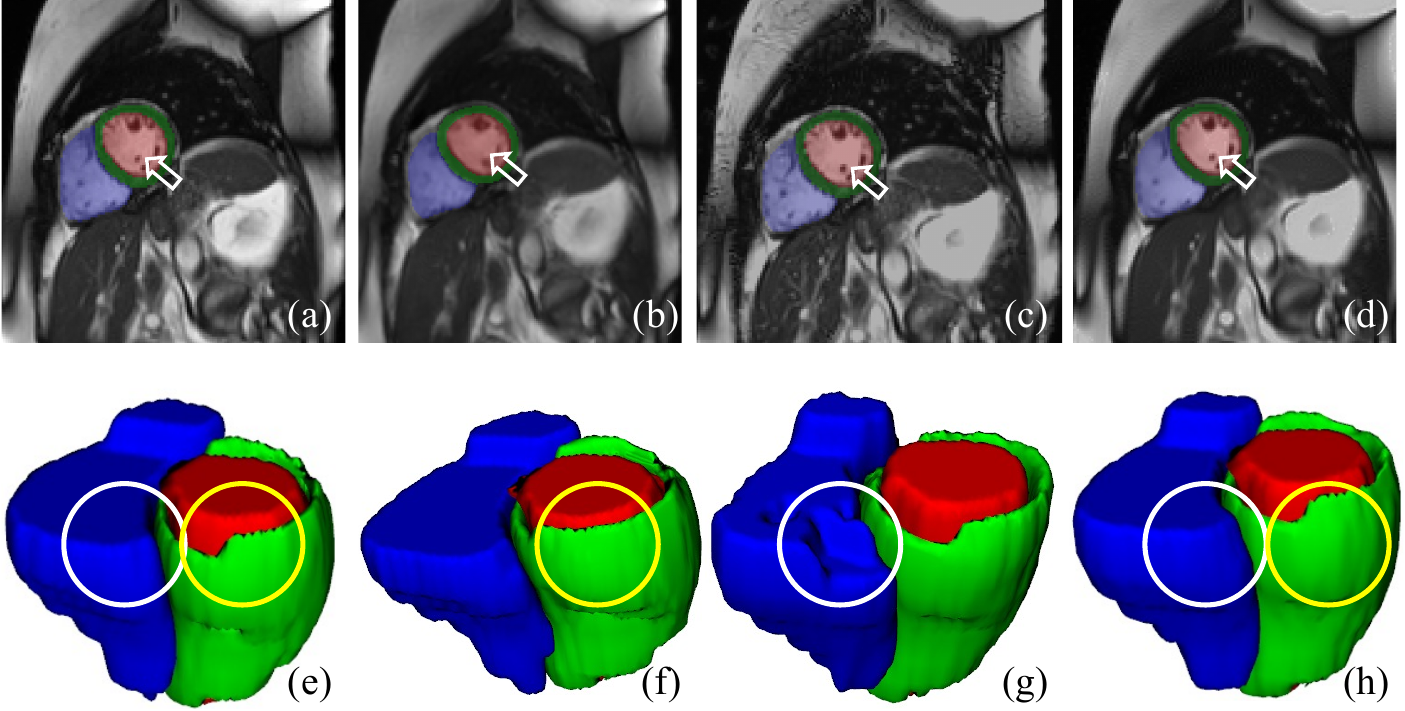}
    \caption{2D and 3D visualization results of the generated images and segmentation. (a,e) original image, (b,f) DeepRecon$_{1k}$, (c,g) DiffAE~\cite{preechakul2022diffusion}, (d,h) our proposed \diffrecon{}.} %
    \label{fig:vis_opti}
\end{figure}
\lastday{We provide peak signal-to-noise ratio (PSNR) and structural similarity index measure (SSIM) \cite{hore2010image} to evaluate the similarity between the generated images and the original images. In addition to image quality assessment,} we want to consider the segmentation performance on the generated images by using a segmentation network trained on the real training data as the evaluator and segment the testing images generated by DeepRecon$_{1k}$, DiffAE which only uses the global semantic latent code as the condition on the DDIM model, and our \diffrecon{} methods. 
The segmentation accuracy of the evaluator on the generated images can be viewed as a quantitative metric to represent the generation quality of the generated data compared to the cMRI data. We compare segmentation obtained based on three  methods against ground truth on the SAX images in \cref{tab:seg-comparison}. The Dice coefficient (DICE), volumetric overlap error (VOE), average surface distance (ASD), Hausdorff distance (HD) and average symmetric surface distance (ASSD) \cite{taha2015metrics} are reported for comparison.

\lastday{Our method achieves a PSNR score of \textbf{30.504} and SSIM score of \textbf{0.982}, which is a significant improvement (35\% increase in SSIM) compared to DeepRecon (PSNR: \textbf{27.684}, SSIM: \textbf{0.724}) with 1k optimization steps.} This indicates that our method generates more realistic image compared to DeepRecon. The segmentation results on the original images in \cref{tab:seg-comparison} provide an upper bound for other results. \diffrecon{} outperforms all other methods in every metric with an 8\% increase in LVM segmentation compared to DiffRecon$_{1k}$
Moreover, by comparing the DiffAE and \diffrecon{}, the introduction of the regional morphology latent code drastically improves the generation results due to the extra information on the shape of LVC, LVM, and RVC. \cref{fig:vis_opti} demonstrates the original image and corresponding synthetic images. The white arrow points towards the presence of cardiac papillary muscles. As indicated in the images, DeepRecon$_{1k}$ (b) cannot effectively recover the information of the papillary muscles from the latent space. However, both diffusion-based (c,d) methods accurately synthesize the information. Our method (d) generates a cleaner image with less artifacts than (c), especially around the LV and RV regions. By comparing the yellow circled area, our method produces image closer to the ground truth compared to DeepRecon$_{1k}$. Also, the white circle in Fig.~\ref{fig:vis_opti} demonstrates the benefits of incorporating regional morphology information. 
Besides, the generative model used in DeepRecon$_{1k}$ needs to be trained for 14 days with additional time to iteratively optimize the latent code for each slice.
Our method uses $4.8$ days for training. Since DDIM inversion does not have test-time optimization as DeepRecon does, \diffrecon{} generates images faster than DeepRecon.
\begin{table}[t]
    \centering
    \caption{Evaluation of 3D volumetric reconstruction from the DICE score of the intersection on each LAX plane against ground truth based on 2D LAX sampled images: mean (standard deviation). Nearest Neighbor, Image-based Linear Interpolation, DeepRecon$_{1k}$ and our \diffrecon{} method are compared.}
    \begin{tabular}{c|c|c|c|c}
    \hline
    Method & Average DICE & 2ch DICE & 3ch DICE & 4ch DICE\\
    \hline
    Nearest Neighbor & 0.780 (0.111) & 0.787 (0.091) & 0.793 (0.105) & 0.766 (0.128)\\
    \hline
    Linear Interpolation & 0.781 (0.080)&0.797 (0.051)& 0.773 (0.070) &0.768 (0.102)\\
    \hline
    DeepRecon$_{1k}$ & 0.817 (0.097)&\textbf{0.848} (0.056) & 0.802 (0.141) &0.797 (0.091)\\
    \hline
\diffrecon{} &\textbf{0.836 (0.052)} & 0.841 \textbf{(0.042)} & \textbf{0.809 (0.069)} & \textbf{0.854 (0.043)}\\ \hline
    \end{tabular}
    \label{tab:recon}
\end{table}
\begin{figure}[t]
    \centering
    \includegraphics[width=0.8\linewidth]{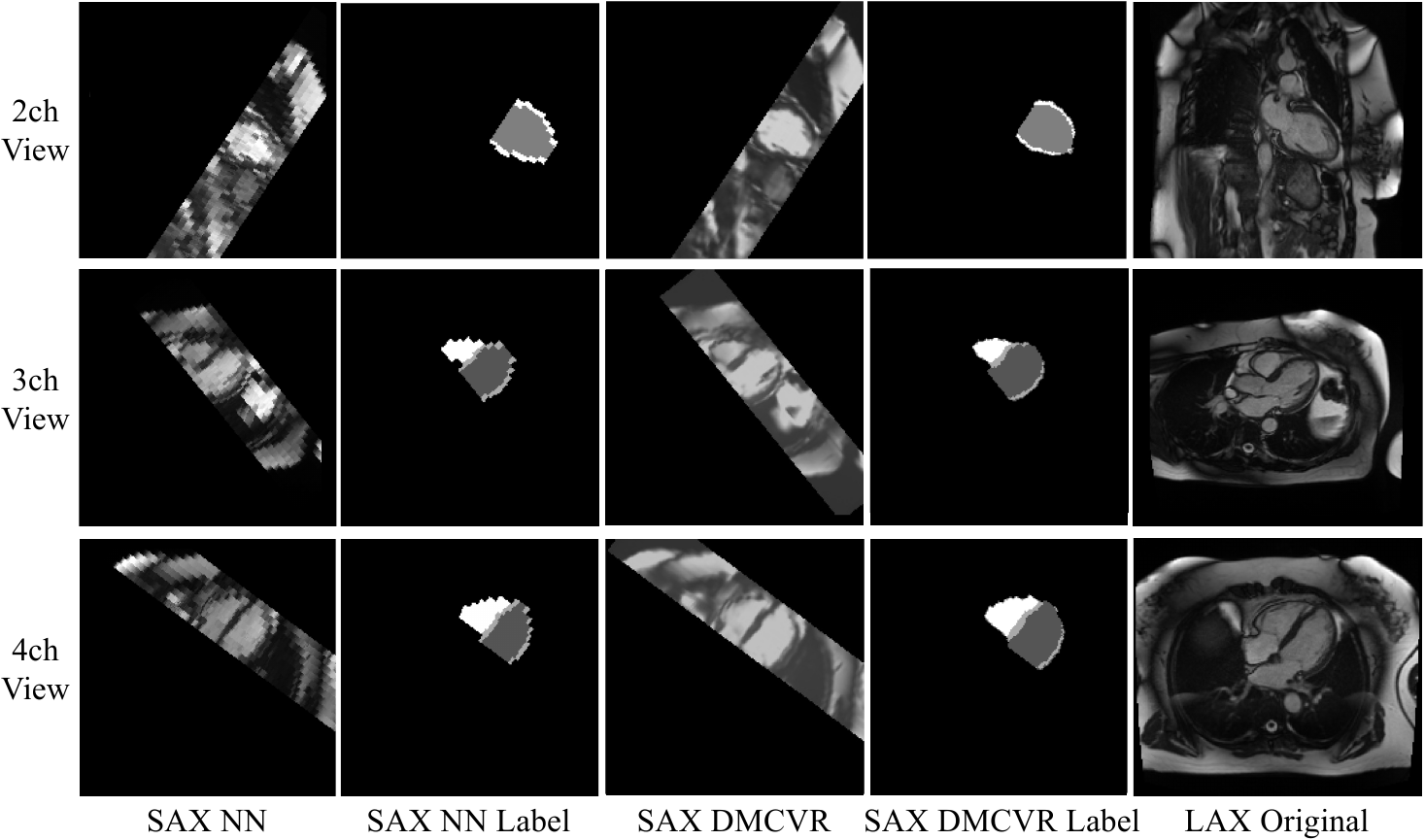}
    \caption{Visual comparison of 3D volumetric reconstruction from SAX images to LAX. Each row from top to bottom are 2ch, 3ch and 4ch images. The column from left to right represents: resampled original images using nearest neighbour (NN), resampled original labels using NN, resampled \diffrecon{} images, resampled \diffrecon{} labels and the corresponding LAX images.}
    \label{fig:recon}
\end{figure}
\subsection{Evaluation of the 3D volume reconstruction quality through latent space interpolation}
In this section, we exploit the relationship between SAX and LAX images and leverage the LAX label to evaluate the volume reconstruction quality. In cardiac MRI, long axis (LAX) slices typically comprise 2-chamber (2ch), 3-chamber (3ch), and 4-chamber (4ch) views. To evaluate the performance of different interpolation methods on LAX slices, we conducted the following experiments: 1) Nearest Neighbor resampling of short-axis (SAX) volume to each LAX view, 2) Image-based Linear Interpolation, 3) DeepRecon$_{1k}$, and 4) our \diffrecon{}. \cref{tab:recon} shows the computed 2D DICE score between the annotation of different LAX views and the intersection between the corresponding LAX plane and 3D reconstructed volume. Our method outperforms other methods in three categories and has only less than 1\% performance degradation compared to DeepRecon$_{1k}$ but with more stable performance. \cref{fig:recon} presents three examples for each LAX view, showing better reconstructed LAX results compared to the original images.

\section{Conclusion}
Integrating analysis of cMRI holds significant clinical importance in understanding and evaluating cardiac function. We propose a diffusion-model-based volume reconstruction method. Our finding shows that through an interpolatable latent space, we are able to improve the spatial resolution and produce meaningful MR images. In the future, we will consider incorporating LAX slices as part of the generation process to help refine the latent space.
\\

\noindent\textbf{Acknowledgement} This research has been partially funded by research grants to D. Metaxas through NSF: IUCRC CARTA 1747778, 2235405, 2212301, 1951890, 2003874, and NIH-5R01HL127661.
%
%
%
\bibliographystyle{splncs04}
\bibliography{paper718}

\end{document}